\documentclass[preprint,superscriptaddress,nofootinbib]{revtex4-2}
\usepackage[utf8]{inputenc}
\usepackage{amsfonts}
\usepackage{amssymb}
\usepackage{amsmath}
\usepackage{amsthm}
\usepackage{mathtools}
\usepackage{mathrsfs}
\usepackage{graphicx}
\usepackage{enumerate}
\usepackage{slashed}
\usepackage{siunitx}

\usepackage[margin=1.1in]{geometry}
\newcommand{\eq}[1]{Eq.~\eqref{eq:#1}}

\newcommand{\fig}[1]{Fig.~\ref{fig:#1}}

\newcommand{\be}{\begin{equation}}
\newcommand{\ee}{\end{equation}}

\begin{document}

\title{Performance envelope of laser wakefield accelerators}
\author{Lance Labun}
\email{lancelabun@tausystems.com}
\affiliation{Tau Systems, Inc., 201 W 5th St, Suite 1100, Austin, TX 78701, USA}
\author{Miguel Gracia-Linares}
\affiliation{Tau Systems, Inc., 201 W 5th St, Suite 1100, Austin, TX 78701, USA}
\author{Ou Z. Labun}
\affiliation{The University of Texas, 2515 Speedway, C1600, Austin, TX 78712 USA}
\author{Stephen V. Milton}
\affiliation{Tau Systems, Inc., 201 W 5th St, Suite 1100, Austin, TX 78701, USA}

\date{20 March 2025}
\preprint{Tau Tech Note 24-01}

\begin{abstract}
Laser wakefield accelerator experiments have made enormous progress over the past $\sim 20$ years, but their promise to revolutionize high-energy particle sources is only beginning to be realized.  To make the next step toward engineering LWFAs for different accelerator outcomes, we need more reliable and quantitative models to predict performance.  Using the data from $>50$ published experiments, we estimate scalings and the performance envelope.  We compare the observed scalings with several models in the literature.  We find that the total beam energy (centroid energy times beam charge) scales almost linearly with laser energy, supporting the value of investment in progressively higher energy driver lasers. The dataset includes pulse durations from 8 to 160 fs, but only laser wavelengths of 800 nm and 1 \si{\micro\meter}, meaning we could not check proposed wavelength scalings for alternative laser technologies. As a benchmark next-generation case, the observed scalings suggest that achieving a 100-GeV LWFA stage will require a $\gtrsim 30$ PW laser operating at electron density $<10^{17}/$cm$^3$.
\end{abstract}

\maketitle
\section{Introduction}

The 2023 P5 report has re-emphasized the need to transition advanced accelerator technologies from university research to refined development in order to realistically contribute to the design of next generation of high-energy particle accelerators \cite{asai2024exploring,geddes24aac}.  Industrial research and development will also benefit from novel accelerator concepts, especially those such as laser wakefield acceleration that promise to make ultra high-energy particle sources more compact, less expensive and thereby more accessible.  To evaluate the return on investment in different accelerator designs and component technologies, we need to consider optimization of the novel accelerator concept in terms of energy efficiency and cost for performance.  We focus on the accelerator performance potential of laser wakefield accelerators, which have been proven to provide 8 GeV, 5 pC \cite{Gonsalves2019} and 10 GeV, $>100$ pC class electron beams \cite{aniculaesei2024acceleration} and promise to improve further with continuing advances and investment in laser technology  \cite{tausystems}.

To design laser wakefield accelerators (LWFAs) according to electron beam requirements, we need a quantitative model relating the laser and plasma inputs to the electron beam outcomes.  Experimentalists have tested a variety of plasma designs from unmodified gas jets to multi-inlet variable-length flow cells to gas capillaries with pre-heating.  The range of laser systems is somewhat narrower, using exclusively lasers with near 1-micron wavelength, and pulse energies ranging from 26 mJ to 130 J (power from 1.7 to 940 TW).  From such a range of experiments, we can reasonably expect to identify some patterns or most promising approaches to accelerator design by comparing various performance metrics.

The number of reported experiments is $>50$, with many reporting a large number of shots and some reporting parameter scans.  This dataset offers a large enough sample to investigate statistically.  While the dataset is larger than is convenient to enumerate by hand, a human can reasonably check the sanity of the data.  The emergence of large language models has highlighted the potential of generative artificial intelligence (AI) models in digesting and summarizing large bodies of text.  We take this opportunity to explore the usefulness of AI in extracting data from heterogeneous sources.  

\section{Metrics}

We quantify accelerator performance by several widely-measured metrics, listed in Table \ref{tab:metrics}.  When discussing these metrics, we have in mind the simplest model of a beam from an accelerator: an energy distribution that is nonzero only over a narrow finite interval around the centroid energy.  Many LWFA experiments fail to realize this ideal model, with energy spreads of several percent or more and/or long tails in the energy distribution, especially at lower energy.  A few display multiple bunches or beamlets.  However, the usable/useful components of these beams will be in general a narrow energy slice where the spectral charge density $dQ/dE$ is highest.  For this reason, we take the nominally reported beam energy and beam charge, which in most cases refers to a peaked feature in the spectrum.  We elaborate on these definitions and caveats in this section.

\begin{table}[h!]
    \centering
    \begin{tabular}{c|c|c|c}
symbol  & description & units & Frac. extracted\\ 
\hline
 $E_e$  & centroid energy  & MeV,GeV & 1 \\ 
 $Q_b$ & charge & pC,nC & 0.72  \\
 $E_b$ & total beam energy & mJ & 0.72 \\
 $\Delta\theta$ & divergence & mrad & 0.60 \\
 $\Delta E/E$ & beam energy spread & $\%$ & 0.72\\
 $\delta E/E$ & shot-to-shot beam energy variance & $\%$ & 0.40\\
 $\delta Q/Q$ & shot-to-shot beam charge variance & $\%$ & 0.23\\
 \hline
   \end{tabular}
    \caption{Summary of electron beam performance metrics.  The last column gives the fraction of papers from which Elicit successfully extracted the number.}
    \label{tab:metrics}
\end{table}

The most basic is the characteristic electron energy, $E_e=\gamma m_ec^2$, or equivalently the characteristic electron Lorentz factor $\gamma=\sqrt{1+\vec{p}_e^2/m_e^2}$.  $E_e$ includes both the rest mass $m_ec^2$ and relativistic kinetic energy $(\gamma-1)m_ec^2$. LWFA experimental publications do not have a standard definition, but due to the nature of the diagnostic, a common definition in practice is the centroid of the electron distribution as recorded by the spectrometer screen.  That is, the reported energy is the deconvolved energy where the peak intensity on the spectrometer screen is observed.  In some cases, a local maximum intensity at the highest beam energy is selected to be reported rather than the global maximum, especially when multiple peaks appear.  The deconvolution should account for the magnetic field distribution in the spectrometer (to varying degrees of accuracy) and sometimes corrects for the electron beam pointing but does not deconvolve the phase space structure of the beam.  Thus beam divergence in the spectrometer's dispersion plane contributes to uncertainty in the centroid energy.  
LWFA beams frequently display a long low-energy tail arising from varying amounts of injection after the primary injection event.  Not all facilities observe the tail due to filtering by magnetic beamline elements: most electrons with energy significantly less than the design energy will be bent too much by any dipoles (for dispersion to the spectrometer) or quadrupoles (for focusing) and exit the beamline\cite{brown1984first} rather than reach the spectrometer.

This long low-energy tail presents difficulties for defining the second important metric, the electron beam charge, $Q_b$.  The ``best'' LWFA outcomes show a few-percent-width peak near the endpoint of the spectrum and much lower spectral charge density ($<10\%$ the peak) outside this peak. However, articles report charge in a variety of ways, ranging from charge ``in'' the peak, with ``in'' undefined, to all charge recorded on the spectrometer.  We do not attempt to sanitize these inputs to the database, because it would require re-analyzing incomplete datasets, eg $dQ/dE$ lineouts in published figures.  This inconsistency in reporting likely introduces noise into our analysis. 

A useful performance metric for overall accelerator performance is the total energy in the beam,
\begin{align}
    E_b= \int E\frac{dQ}{dE} dE,
\end{align}
with the left-hand side measured in joules when the particle energy is measured in eV and spectral charge density in C/eV.  The integral should be defined consistently with the beam charge, and our evaluation of $E_b$ thus inherits the difficulties with the charge data reporting just described.  Again thinking of the ``best'' LWFA outcomes as a narrow peak near the endpoint of the spectrum, we characterize this feature in the beam by its energy $E_e$ and first moment $\Delta E/E$, allowing a simple parameterization of the integral:
\begin{align}\label{eq:defnEb}
    E_b\simeq  E_eQ_b\left(1+c_\rho \frac{\Delta E}{E}\right)
\end{align}
In the limit of an infinitely narrow peak $\Delta E/E\to 0$, $dQ/dE\to Q_b\delta(E-E_e)$ where $\delta(x)$ is the Dirac delta distribution.  For a small nonzero width $1\gg\Delta E/E>0$, the integral can be corrected by a term linear in $\Delta E/E$ with $c_\rho$ a numerical coefficient of order 1, depending on the shape of the distribution around the peak.  For example, for a gaussian, $c_\rho=(2\ln 2)^{-1}$.  We emphasize that $E_b$ represents the energy in the reported beam, which in practice is usually a post-selected peak feature in the spectrum.

\begin{table}[h!]
    \centering
    \begin{tabular}{c|c|c|c}
 symbol & description & units & Frac. extracted\\ \hline
 $E_\ell$  & laser pulse energy  & J & 0.99\\
 $\tau_\ell$ & pulse duration & fs &  1 \\
 $P_\ell$ & laser peak power & TW,PW & 0.97\\
 $w_0$ & laser spot size & $\mu$m & 0.60 \\
 $a_0$ & laser normalized peak amplitude & -- & 0.62\\
 $n_e$ & plasma density & /cm$^3$ & 0.94\\
 -- & injection type & -- & 0.99 \\
 \hline
   \end{tabular}
    \caption{Summary of accelerator control parameters.}
    \label{tab:accelparams}
\end{table}

We also classify LWFA experiments by their injection method, including self, shock, downramp, ionization and nanoparticle.  These different methods are not always carefully distinguished.  For instance, a wire or blade introduced in the gas stream can produce either shock or downramp injection, roughly distinguished by the magnitude of the density gradient observed on the beam axis.  For this reason, we have tried to consistently label to downramp injection where variable gas pressures and/or channel geometry control the density profile and shock injection where an obstacle is introduced in the gas flow to induce a shock-like feature in the density profile.

\section{Models}

Models are provided in the literature to estimate beam energy and beam charge without defining what these quantities mean with respect to the distribution.  Since there is usually less than a factor of 2 between the endpoint of the spectrum (the maximum recorded energy) and the nominal beam energy as reported and described above, we can assume that these models intend to predict the maximum or ideal beam energy and expect that they overestimate the nominal beam energy by a factor close to 1. 

The models are derived from 1-dimensional analytic solutions for wakefields \cite{esarey2009physics} and 3-dimensional simulations \cite{lu2007generating}.  Predictions of beam energy are constructed from estimates of unmeasured quantities such as the acceleration length, bubble size, electrostatic field in the bubble. As such, different findings for the scaling of these intermediate quantities combine to give significantly different scalings for the observable $E_e$, which are summarized in Table 1 of Ref \cite{lu2007generating}.

For baseline comparisons, we use two models. The ``matched'' model of \cite{lu2007generating} assumes that we should set the vacuum laser spot size $w_0$ equal the expected equilibrium bubble radius.  With this constraint, the laser power, laser wavelength and electron density determine the expected electron energy,
\begin{align}\label{eq:luEemodel}
\frac{E_e}{\rm GeV} 
&\simeq 1.7\left(\frac{P}{100 \rm TW}\right)^{1/3}\left(\frac{n_e}{10^{18}\,/\mathrm{cm}^3}\right)^{-2/3}\left(\frac{\lambda_\ell}{\rm 0.8\,\mu m}\right)^{-4/3}
\end{align}
as given in Eq 6 of Ref. \cite{lu2007generating}.\footnote{In the dimensionless units of the paper, Fig 3 of Ref. \cite{lu2007generating} shows an almost linear electric field across the majority of the bubble, 
$\frac{eE_z(x_v)}{m\omega_p}=\frac{n_e}{2n_{cr}}x_v\omega_\ell$ where $x_v=x-vt$ is the comoving coordinate and the prefactor is the fit to the line given in the paper.  Setting $x_v=R$, we obtain 
$\frac{eE_{\rm LW}}{m\omega_p}=\frac{E_{z,max}}{E_0}=\sqrt{a_0}\frac{\omega_p}{\omega_\ell}=\sqrt{a_0\frac{n_e}{n_{cr}}}$.  This estimate is consistent with bubble radius and electric field seen in Fig 3 of Ref. \cite{lu2007generating}, but does not agree with the assertion above Eq 6 in that manuscript.  However the resulting equation significantly under-predicts electron energy.}  Seeing that few experiments explicitly aim to match bubble radius and spot size, we construct a simpler model by neglecting nonlinear corrections to the bubble radius.  The energy gain is still estimated as half the maximum electric field times the dephasing length,
\begin{align}\label{eq:naiveEemodel}
\frac{E_e}{\rm GeV} 
&\simeq 1.9\left(\frac{n_e}{10^{18}{\rm /cm}^3}
\right)^{-1}\left(\frac{\lambda_\ell}{\rm 0.8\,\mu m}\right)^{-2},
\end{align}
This ``naive'' model does not explicitly depend on laser power and instead reproduces a simple inverse relationship between electron energy and plasma density.  Dependence on laser power is instead implicit in requirements for self-focusing and depletion length being longer than dephasing length.  A third scaling, defined with RF accelerator engineering in mind, is derived from the observation that the power draw of a circuit element is proportional to the voltage drop squared, and therefore the acceleration gradient in the RF cavity scales with the square root of the input power, $\Delta V\propto P_{\rm in}^{1/2}$.  

For beam charge, Ref. \cite{lu2007generating} gives a scaling with laser power and wavelength,
\begin{align}\label{eq:luQmodel}
    \frac{Q}{\rm pC}\simeq 400\left(\frac{P_\ell}{100 \mathrm{TW}}\right)^{1/2}\left(\frac{\lambda_\ell}{0.8\,\si{\micro\meter}}\right).
\end{align}
However the reasoning apparently uses the questionable assumption that all of the electrostatic field energy in the wake is transferred to the beam.  Simulations by Ref. \cite{Brunetti2022} exhibit a distinctly sub-linear scaling of beam charge with laser power\footnote{In that manuscript, $Q$ is plotted versus $E_\ell$ but the pulse duration is fixed to 500 fs across all simulations, so one can also read off how charge varies with peak power.}, but no fit or trend lines were given to compare to the $P^{1/2}$ scaling.  


\section{Methods}
We use the AI tool Elicit \cite{elicit} in its ``data extraction'' mode to quickly extract important parameters and outcomes from Refs
\cite{mangles2004monoenergetic,geddes2004high,faure2004laser,Leemans2006,Osterhoff2008,Brunetti2010,Schmid2010,Cipiccia2011,Weingartner2012,Albert2013,Buck2013,Kim2013,wang2013quasi,Golovin2015,Guillaume2015,Mirzaie2015,Thaury2015,Wang2016,Couperus2017,Swanson2017,Ekerfelt2017,li2017generation,Delbos2018,Irman2018,Tsai2018,Aniculaesei2019upramp,aniculaesei2019proof,Gonsalves2019,gilljohann2019direct,wu2019energy,Shalloo2020,Maier2020,gotzfried2020,jalas2021bayesian,ke2021optimization,Ke2021neargev,Wang2021,Foerster2022,Miao2022,Zhu2023,Pak2023,Liu2023,Lei2023,Pandari2023,aniculaesei2024acceleration,Lazzarini2024,Nam2024,Shrock2024}.  We verified the extracted data two ways.  First, a human read and plotted the tabulated data to identify obvious outliers that might represent incorrectly extracted numbers. As a language model, Elicit highlights the inconsistency in terminology and definitions even within the wakefield accelerator community: we had to sanitize Elicit's output to ensure metrics -- even as basic as ``electron energy'' and ``laser energy'' -- are correctly identified.  Results exhibited in table format, with standard terminology and common units, were most consistently extracted correctly.  Most extracted parameters and results had error rates in the $10-20\%$ range, with difficulties concentrated on reported charge and papers that reported multiple results without tabulating outcomes versus inputs. Notably, qualitative distinctions, such as injection type, were extracted without error \cite{spillias2024evaluating}; the limitation lay in the inconsistency of different authors' definitions.  Second, we ran the fits and analysis on a smaller set ($N=26$) of articles that was analyzed with Elicit and compared the resulting trends to a human-extracted set ($N=40$) of articles that overlapped.  The results of fits were consistent within the uncertainty of the fits, showing that errors in the extraction are comparable to the scatter in the data. However, the biggest limitation in this analysis was the incompleteness of many reports.  A significant fraction of experiments did not measure the beam charge, which means that total beam energy \eq{defnEb} could not be computed.

The majority of articles before 2016 report metrics for single shots; a small majority of the articles published 2016 and later report metrics averaged over multiple shots, usually 10-50. More recent manuscripts report 10s of shots, and we reduce the outcomes to averages and variances where they are not already binned and averaged (e.g. \cite{aniculaesei2019proof}).  In several cases (e.g. \cite{Couperus2017,Swanson2017}), this required human intervention because variances were only reported in figures.  Fits are weighted by the number of shots, crediting those experiments that reported reproducibility.  One outstanding exception, Ref. \cite{Maier2020}, reports averages over $\sim 10^5$ shots, 2 orders of magnitude more than the rest of the literature combined.  To prevent Ref \cite{Maier2020} from constraining the fits, we artificially reduced its relative weight by a factor of $100$.

Our primary objective is to predict plasma accelerator performance in order to guide design of the laser and plasma channel.  Looking forward to the potential and future development of LWFAs as an accelerator technology, we want the scaling of the metric $y$ with the input $x$. For this reason, we fit the data with a power law 
\begin{align}
    y = Cx^\beta\,
\end{align}
using the least-squares metric.  Models of wakefield output notably scale with fractional powers of the input, and the dataset is large enough to distinguish the power laws with statistical significance.


\section{Results}

\begin{figure}
    \includegraphics[width=0.48\textwidth]{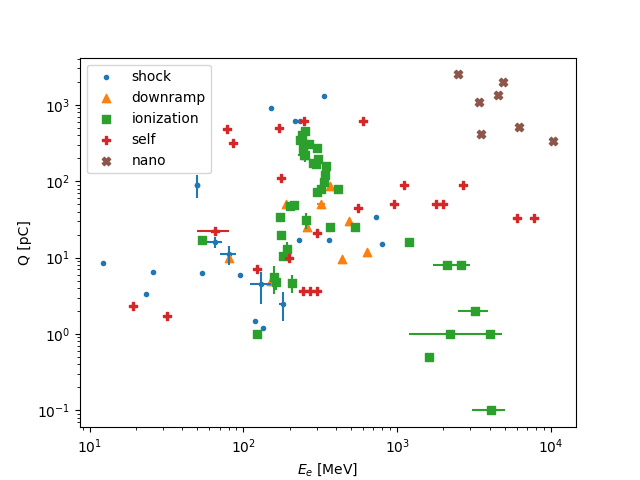}
    \includegraphics[width=0.48\textwidth]{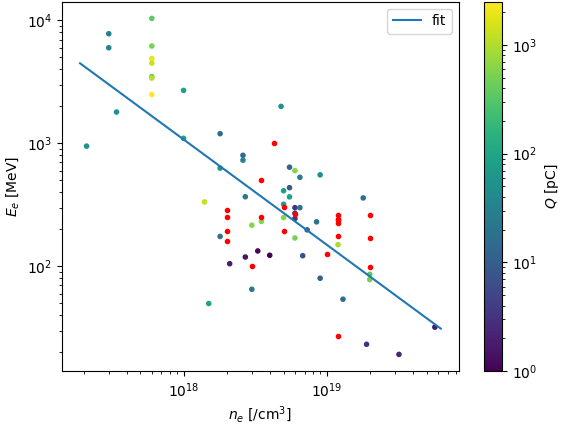}

    \caption{Left: Scatter plot of reported beam charge versus beam energy.  Symbols indicate injection mechanism.  Points representing averages over many shots are given with error bars showing the reported shot-to-shot variance. Right: Electron energy versus plasma electron density.    Blue line is a least-squares fit for each case with the model parameters given in text.  Points are colored by beam charge with red points signifying no charge reported.  \label{fig:QvEandEvsn} }
\end{figure}

To introduce the dataset, Figure \ref{fig:QvEandEvsn} is a scatter plot of electron energy versus beam charge, complementing and updating Figure 3 of Ref. \cite{downer2018diagnostics} with recent experiments.  Not all experiments analyzed could be included in \fig{QvEandEvsn} due to omission of charge data.  

The scatter suggests that only nanoparticle injection (labeled 'nano') might be exceptional in combining high charge and high energy beams.  However, all nanoparticle injection events are from a single experiment on the Texas Petawatt \cite{aniculaesei2024acceleration} (an earlier experiment, Ref. \cite{aniculaesei2019proof}, reported only uncalibrated, relative charge), and we will see later that this performance aligns with the trends in electron energy and beam energy set by experiments on lower-energy lasers.  One might also notice that only shock injection, self-injection and nanoparticle injection have provided beams with $Q\gtrsim 100$ pC, despite the expectation that ionization injection should enhance beam charge.  The error bars show shot-to-shot variances in electron energy and beam charge where reported.  Note some error bars are obscured by the markers: some experiments claim few percent-level shot-to-shot variance in electron energy \cite{Couperus2017,Pandari2023}, reflecting perhaps both stability improvements coming with higher repetition rate operations as well as possible post-selection of reported shot sets.

\subsection{Electron energy vs plasma density}

Since decreasing the plasma density decreases the field strength but increases the acceleration length faster, it is widely expected that higher electron energy is achieved with lower plasma density \cite{lu2007generating, schroeder2010physics, siders2019wavelength}.  Because self-guiding at lower density requires higher laser power, the dependence of the energy gain on the density is a crucial question for laser development.  On the other hand, guiding with a preformed low-density channel may help relax laser power required to achieve larger acceleration lengths. 


In the data, we find that
\begin{align}\label{eq:Eenefit}
    \frac{E_e}{\rm GeV}\simeq 0.83 \left(\frac{n_e}{10^{18}/{\rm cm}^3}\right)^{-0.86\pm 0.09},
\end{align}
with order of magnitude uncertainty in the prefactor not written out.
The power law is close to but significantly different from the naive expectation of $n_e^{-1}$. 
Binning by density and fitting only the top 10\% of $E_e$ suggests that the scaling of the maximum achievable energy at each density scales more strongly, close to $n_e^{-1}$. One experiment using a guiding channel collected enough data for different channel densities to see evidence of a $n_e^{-1}$ trend in energy, though the statistics are too weak to conclude anything \cite{Miao2022}.  


\subsection{Electron energy vs laser power}

Predictions for dependence of the electron energy on laser power vary widely: the matched regime predicts a relatively weak power law $P^{1/3}$, a simple model of an RF cavity suggests $P^{1/2}$, and an eyeball estimate on the available data given by \cite{wenz2020physics} suggests an upper bound linear in $P$: $E_e[{\rm MeV}]\leq 10P_\ell[{\rm TW}]$.  A fit to the data shows that the average performance lies in between, scaling as
\begin{align}
     \frac{E_e}{\rm MeV}\simeq 10\left(\frac{P_\ell}{\mathrm{TW}} \right)^{0.81\pm 0.06},
\end{align}
with about $30\%$ uncertainty in the prefactor, which otherwise coincides exactly with the coefficient given by Ref. \cite{wenz2020physics}.

\begin{figure}
\includegraphics[width=0.48\textwidth]{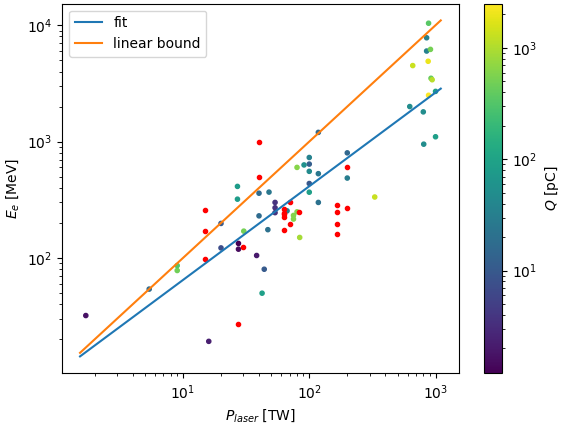}
\includegraphics[width=0.48\textwidth]{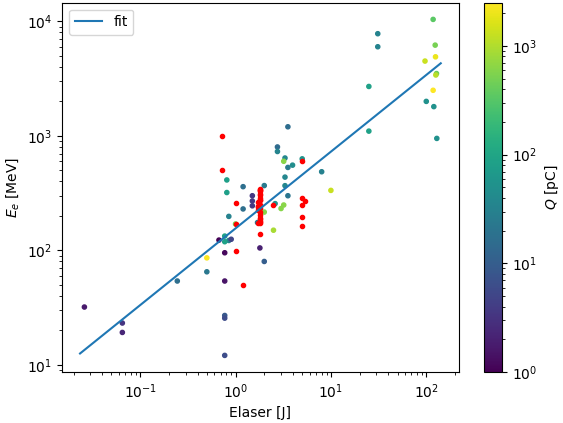}
\caption{Left: Electron energy versus laser peak power.  Right: Electron energy versus laser pulse energy.  Blue line is a least-squares fit, given in \eq{Eenefit}.  Orange line at left is the linear bound offered by Ref. \cite{wenz2020physics}. Points are colored by beam charge with red points signifying no charge reported.  \label{fig:Eevslaser} }
\end{figure}  

\subsection{Electron energy vs laser energy}

There are no models predicting a direct relationship between laser energy and electron energy.  Instead, we expect this to reflect engineering practicalities that arise with the construction of higher energy and higher power lasers.  Since most current high power laser systems utilize Ti:sapphire as a gain medium, peak power and laser energy are closely related by the current typical range of pulse duration (25-35 fs).  However, there are notable exceptions to this rule in the dataset: Refs. \cite{geddes2004high,Albert2013,Kim2013,Swanson2017,Shrock2024} used Ti:sapphire systems with slightly longer pulse durations ($> 45$ fs), and Refs. \cite{aniculaesei2024acceleration,wang2013quasi} used the Texas Petawatt $\sim 120$ J, $130$ fs laser.  The differences of these experiments from the more common peak power-pulse energy relationship suffice to affect the fit.  We find that electron energy scales with laser energy close to a 2/3 power:
\begin{align}
     \frac{E_e}{\rm MeV}\simeq 170\left(\frac{E_\ell}{\mathrm{J}} \right)^{0.65\pm 0.04},
\end{align}
with less uncertainty in both fit parameters and somewhat smaller residuals.  In this sense, laser pulse energy is more effective than laser peak power at predicting the accelerator performance.  We note that the linear bound suggested by Ref. \cite{wenz2020physics} does not match well here, possibly due to the addition of post-2020 experiments at both high \cite{aniculaesei2024acceleration} and low \cite{Lazzarini2024} laser energy.

\subsection{Energy spread vs electron energy}

\begin{figure}[t!]
\includegraphics[width=0.48\textwidth]{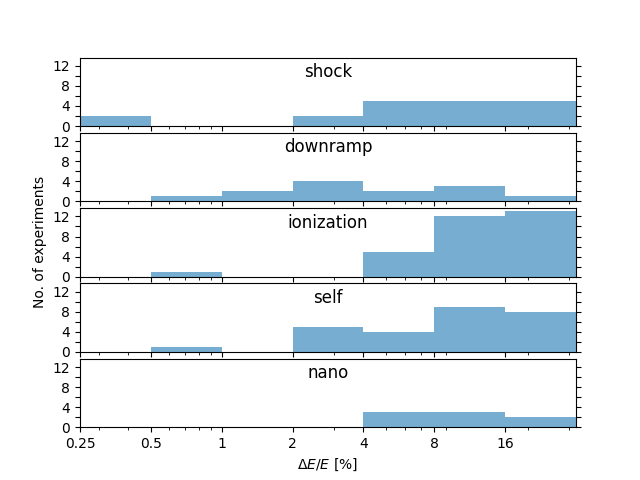}
\includegraphics[width=0.48\textwidth]{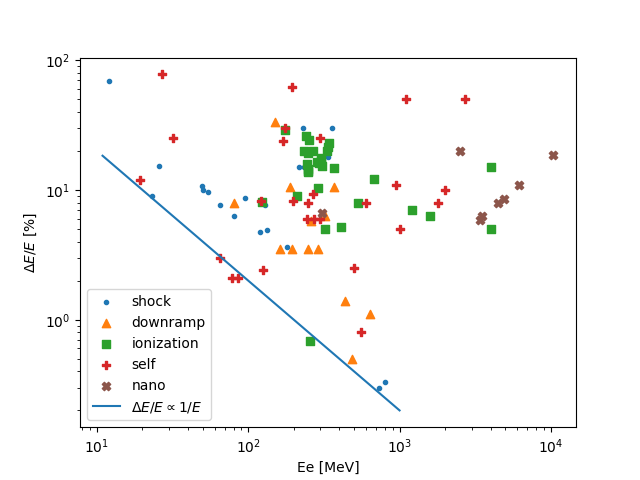}
\caption{Left: Histogram of number of experiments achieving a given energy spread, separated by injection mechanism.  Right: Relative energy spread $\Delta E/E_e$ versus electron energy $E_e=\gamma m_ec^2$.  Blue line is an estimated lower bound where the absolute energy spread is 2 MeV.  \label{fig:dEplots} }
\end{figure}

There are no models predicting a direct relationship between energy spread and laser or plasma input parameters.  The general expectation is that injection mechanism, especially sharp density transitions \cite{Schmid2010,Buck2013,suk2001plasma,tomassini2003production} or colliding pulses \cite{faure2006controlled,kotaki2008improvement}, can tune the energy spread.  A few works have demonstrated some level of systematic control of energy spread by varying parameters of the density profile \cite{Swanson2017,Tsai2018,wu2019energy,ke2021optimization}.  In contrast, ionization injection is  expected to lead to high-energy spread bunches, which may be an acceptable side effect when the goal is maximizing beam charge.  In the ensemble of experiments \fig{dEplots} (left), we see circumstantial evidence for the latter conclusion, but less evidence to favor any particular injection mechanism.  If reducing energy spread is a priority, one must still study individual cases where exceptional energy spread has been achieved (e.g. \cite{jalas2021bayesian,Ke2021neargev,Wang2021}) and where energy spread has been successfully varied as a function of input parameters.

On the other hand, a lower bound on the measured absolute energy spread arises from the finite time interval over which injection occurs.  The difference in time and longitudinal position of the wake between the first and last injection events causes a difference in acceleration length, translating into energy spread.  The minimum time interval for injection is set by plasma dynamics.  Injection into the moving wake potential requires crossing the phase space separatrix between trapped, co-moving trajectories and non-trapped trajectories \cite{esarey1995trapping}.  Self-injection, shock injection and downramp injection all achieve this by perturbing the wake via the plasma.  The shortest distance over which the wakefield can respond is a few skin depths $c/\omega_{\rm pl}$; for example, after an ideal, instantaneous density step, the plasma wave relaxes to its new wavelength after at least a few skin depths.  In contrast, ionization injection works by releasing free electrons in the trapping region, but is difficult to localize to regions of the plasma smaller than hydrodynamic length scales, which is why we generally expect and observe higher energy spread. 

Estimating the resulting energy spread as $\Delta E_e\simeq \overline{eE_z}\Delta z_{\rm inj}$ where $\overline{eE_z}$ is the average accelerating gradient and $\Delta z_{\rm inj}$ in the spatial distance between first and last injection, we find two features: (1) $\Delta E_e$ is independent of plasma density in agreement with the data, and (2) for typical values (not shown), $\Delta E_e\simeq $ a few times $m_e$.  The trend line in \fig{dEplots} (right) is for an absolute energy spread $\Delta E_e= 2$ MeV.  Note that Ref. \cite{jalas2021bayesian}, represented by the green square near the $E^{-1}$ fit line and which uses a combination of ionization and downramp injection.  The set of 9 blue, shock injection points falling near a $E^{-1}$ line are all from Ref. \cite{Buck2013}, showing that the scaling holds approximately within shock injection and within a single experimental setup as plasma parameters varied.  The absolute energy spread averaged over configurations in Ref. \cite{Buck2013} was 6 MeV.
 
Other effects could place lower bounds on the energy spread but either place smaller (less limiting) lower bounds or vary more from experiment to experiment.  Finite injection interval would be considered an ``intrinsic'' source of energy spread, arising from the physical process initiating the beam.  Another intrinsic source is variance in transverse momentum, but this is at least two orders of magnitude smaller and therefore not a limiting factor.  Space charge effects are strongly suppressed $\propto 1/\gamma^2$ for LWFAs since the initial gradient is so high.  Other dynamical sources, such as differences in the acceleration gradient between the head and tail of the beam, can be suppressed if the bunch length is much less than the plasma wavelength and need not always increase the energy spread.

\subsection{Beam energy vs laser energy}

\begin{figure}
\includegraphics[width=0.48\textwidth]{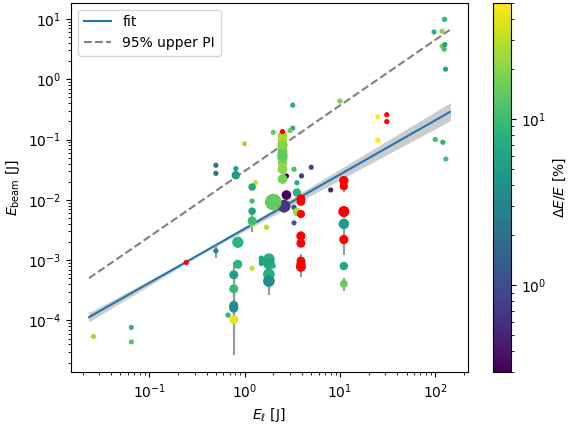}
\caption{Beam energy \eq{defnEb} versus laser energy. Blue line is a least-squares fit, given in \eq{Ebfit}, and the dashed line is the $95\%$ upper prediction interval, given in \eq{EbupperPI}.   Points are colored by energy spread with red points signifying no energy spread reported.  Point size shows the relative number of shots contributing to the measurement.  Grey (vertical) lines exhibit the shot-to-shot variance where available.  \label{fig:Ebvslaser} }
\end{figure}

An essential metric that has not been studied before is the beam energy, as defined in \eq{defnEb}.  For average performance, we find
\begin{align}\label{eq:Ebfit}
    \frac{E_b}{\rm mJ} \simeq (3.3\pm 0.2)\left(\frac{E_\ell}{\rm J}\right)^{0.9\pm 0.05} \qquad\mathrm{(average)},
\end{align}
showing that $\simeq 0.3\%$ efficiency is typical near the common operating point of a few joule per pulse.  The collection of points in the top right of Figure \ref{fig:Ebvslaser} is from Ref. \cite{aniculaesei2024acceleration}, suggesting that  $0.1-1$ nC beam charge at $E_e\simeq 10$ GeV should be the expected outcome of a 100 J-class laser, though in that case it was only achieved with addition of nanoparticles.

The scaling is notably nearly linear with laser energy, implying that the efficiency $E_b/E_{\ell}$ is only very weakly dependent on laser energy,
\begin{align}
    \frac{E_b}{E_\ell}\simeq 0.003\left(\frac{E_\ell}{\rm J}\right)^{-0.1} \qquad\mathrm{(average)},
\end{align}
We have verified that this insensitivity carries over to laser power.  This result is especially good news for advanced accelerator development, because it implies that the plasma can generally be adapted to the laser, and investment in high-power laser technology, especially delivering $\geq 100$ J pulses at high repetition rate, directly transfers to accelerator performance without diminishing returns.

Note that each point in the dataset represents a potentially usable quasi-monoenergetic feature in the observed spectrum.  The complete accelerated bunch may consist of multiple beamlets or a broader, flatter background distribution in addition to the quasi-monoenergetic feature, such that a significantly greater fraction of laser energy is usually transferred to high-energy electrons.  However these extraneous features are generally not helpful for either collider or secondary particle source applications.  We consider efficiency into the usable component the most relevant metric for design and development planning.

Finally to estimate the performance envelope, i.e. the maximum possible beam energy for a given laser, we construct the $95\%$ upper prediction interval from the fits and their variances, 
\begin{align}\label{eq:EbupperPI}
    \frac{E_b}{\rm mJ} \simeq (30\begin{smallmatrix}+12 \\ -9 \end{smallmatrix})\left(\frac{E_\ell}{\rm J}\right)^{1.09\pm 0.14} \qquad \mathrm{(95\%~upper~PI)}.
\end{align}
In short, well-designed LWFAs can achieve $3\%$ conversion efficiency, with no significant penalty as the laser energy is increased.  This envelope may be marginally improved as the use of active feedback and optimization methods \cite{Shalloo2020,jalas2021bayesian,irshad2024pareto} becomes more common in LWFA experiments, but we expect that the dataset here is large enough that the present interval is a very good estimate of the physical limits.  

On the other hand, total beam energy is not necessarily the target of all optimizations, since some observables (such as betatron radiation) scale differently with respect to particle energy and beam charge and other departures of the accelerator from its most-efficient operating point may enhance the desired objective (see e.g. \cite{Shalloo2020})

\section{Conclusions}

By examining the ensemble of reported LWFA experiments, we have derived phenomenological scaling relationships for accelerator performance as a function general laser and plasma parameters.  Seeing that most experiments do not restrict their operating point to the predicted ``matched'' regime of \cite{lu2007generating}, it is not surprising that the ``matched'' regime scalings are not reproduced.  Nevertheless, departure of the observed scalings from models demands attention from theorists, in order to build more accurate and comprehensive models for LWFA design.

The discovered scalings imply that a 100 GeV LWFA stage can be easily achieved with an 87 PW, 18 kJ laser (corresponding to 210 fs) operating at $n_e<10^{17}/\mathrm{cm}^3$ plasma density.  A 1 TeV LWFA stage would call for an exawatt-scale, 600-kJ laser.  These combinations of energy and peak power leave open the possibilities that Nd:glass remains and Tm:YLF becomes a competitive laser technology for ultra-high energy LWFA accelerators \cite{siders2019wavelength}.  Since these scalings are obtained from average performance, we expect that a 3-4$\times$ smaller system could suffice, fully utilizing lessons from the literature on optimizing the accelerator output.  

\begin{figure}
\includegraphics[width=0.48\textwidth]{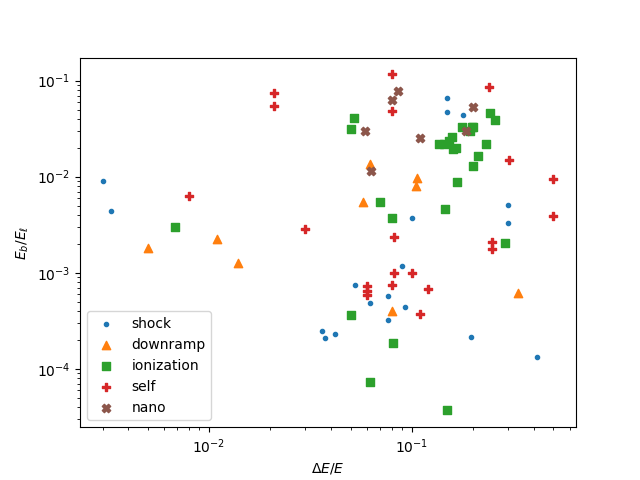}
\caption{ Energy efficiency $E_b/E_\ell$ versus relative energy spread of the resulting bunch $\Delta E/E$.  Symbols show injection type. \label{fig:etavsdE} }
\end{figure}

On the other hand, no experiment has yet produced a beam with narrow enough energy spread at a reasonable efficiency to demonstrate a step toward supplanting RF technology for high-energy particle colliders.  As shown in \fig{etavsdE}, the top-left corner $\Delta E/E<1\%$ and $E_b/E_\ell>1\%$ is empty.  The nearest cases are from closely related reports, Refs. \cite{ke2021optimization} and \cite{Ke2021neargev}, which demonstrated $E_e>700$ MeV, 15-30 pC bunches.  Otherwise, the highest efficiency accelerators are associated with larger energy spreads, in most cases provided by ionization or nanoparticle injection.  Fortunately, a large space of non-collider applications that have less stringent efficiency and energy spread requirements stands to benefit from the engineering advantages of the laser-driven high-energy particle accelerator.  The scalings  here provide a first basis for economic estimates for these applications, since construction and operating costs will be driven by laser energy and repetition rate.

\begin{acknowledgments}
O.Z.L. is supported in part by the Sponsored Research Agreement NO. UTAUS-FA00001488 Between THE UNIVERSITY OF TEXAS AT AUSTIN And TAU SYSTEMS INC.
\end{acknowledgments}

\bibliographystyle{apsrev4-2}
\bibliography{lwfa}

\end{document}